\documentclass[10pt]{article}
\usepackage{axodraw}
\usepackage{graphicx}
\usepackage{cite,./mcite}
\usepackage{epsfig}
\usepackage[latin1]{inputenc}

\begin{document}
\title{Multiple Scattering, Underlying Event, and Minimum Bias }
\author{G\"osta Gustafson\protect\footnote{ talk presented at 12th Int. Conf. 
on Elastic and Diffractive Scattering, EDS07, DESY, Hamburg, 21-25 May 2007}\\
Dept. of Theoretical Physics, Lund Univ.\\ 
S\"olveg. 14A, S-22362 Lund, Sweden\\
e-mail: Gosta.Gustafson@thep.lu.se }
\date{}
\maketitle
\begin{flushright}
  \vspace*{-7.5cm}
  LU-TP 07-39\\
 September 2007
  \vspace*{6cm}
\end{flushright}
\begin{abstract}
In this talk I first discuss the experimental evidence for multiple scattering 
and the properties of the underlying event. The extensive analyses by Rick
Field of data from CDF cannot be reconciled with traditional wisdom conserning
multiple collisions and the AGK cutting rules. Data seem to imply some kind of
color recombination or unexpectedly strong effects from pomeron vertices.

I then discuss theoretical ideas concerning the relation between multiple
collisions and unitarity: the AGK rules, IP loops, dipole cascade models 
and diffraction.
 
\end{abstract}

\section{Experimental overview}
\label{sec:mult}

\subsection{Minijet cross section}

In \emph{collinear factorization} the cross section for a parton-parton subcollision is given by
\begin{equation}
\frac{d\sigma^{subcoll}}{d p_\perp^2} \sim \int dx_1 dx_2 f(x_1, p_\perp^2)
f(x_2, p_\perp^2) \frac{d\hat{\sigma}}{d p_\perp^2}
(\hat{s}=x_1 x_2 s,p_\perp^2).
\label{eq:sigmasubcoll}
\end{equation}
(Note that one hard subcollision corresponds to 2 jets.) The partonic cross section
$d\hat{\sigma}/d p_\perp^2$ behaves like $1/p_\perp^4$ for small $p_\perp$, which means that
a lower cutoff, $p_{\perp min}$, is needed. The total subcollision cross
section is then proportional to
$1/p_{\perp min}^2$, and for $pp$-collisions the subcollision cross section
becomes equal to the total cross section for $p_{\perp min}\approx$ 2.5 GeV at the Tevatron
and $\approx$ 5 GeV at LHC \cite{Sjostrand:2004pf}. Fits to data give $p_{\perp min}
\sim$ 2 GeV at the Tevatron and slowly growing with energy \cite{Sjostrand:1987su}.

In \emph{$k_\perp$-factorization} there is a dynamic cutoff when the momentum
exchange $k_\perp$ is smaller than the virtuality of the two colliding partons, given by
$k_{\perp 1}$ and $k_{\perp 2}$ \cite{Gustafson:1999kh, *Gustafson:2002kz}. 
This approach gives a very  similar effect. Thus we conclude that at high
energies the subcollision cross section is 
much larger than the total inelastic cross section, which means that on average
there must be several hard subcollisions in one event. It was also early suggested
that the increase in $\sigma_{tot}$ is driven by hard parton-parton subcollisions
\cite{Cline:1973kv, *Ellis:1973nb, *Durand:1987yv}.

\subsection{Experimental evidence for multiple collisions}

\subsubsection{Multijet events}

Besides from independent multiple subcollisions, multijet events can also
originate from multiple bremsstrahlung from two colliding partons. If we study
four-jet events the difference between these two types of events is that in
a double parton scattering the four jets balance each other pairwise
in the transverse momentum plane, while such a pairwise balance is not present
in the multiple bremsstrahlung events. The Axial Field Spectrometer at the
ISR proton-proton collider \cite{Akesson:1986iv} studied an "imbalance parameter"
\begin{equation}
J=\frac{1}{2} [(\mathbf{p}_{\perp 1}+\mathbf{p}_{\perp 2})^2 +
(\mathbf{p}_{\perp 3}+\mathbf{p}_{\perp 4})^2],
\label{eq:imbalance}
\end{equation}
and found that there is a significant enhancement of events with small values of $J$,
which thus showed a clear evidence for multiple subcollisions.

Similar, but less clear, results for four-jet events have been observed by the CDF
\cite{Abe:1993rv} and D0 \cite{Abazov:2002mr} experiments at the Tevatron.
A more clear signal for multiple collisions at the Tevatron has instead been seen
in events with three jets + $\gamma$ \cite{Abe:1997xk}. Evidence for multiple
collisions has also been observed in photoproduction by the ZEUS collaboration
at HERA \cite{Gwenlan:2002st}.

\subsubsection{Underlying event}

An inportant question is whether the hard subcollisions are correlated, or if
a high $p_\perp$ event just cooresponds to two jets on top of a minimum bias event.
If the subcollisions are uncorrelated the probability, $P(n)$, for having $n$ subcollisions
should be described by a Poisson distribution. This implies that
\begin{equation}
P(2)=\frac{1}{2} P(1)^2.
\label{eq:P2}
\end{equation}
Here the factor 1/2 is compensating for double counting. Expressed in the cross sections
$\sigma_n=P(n) \sigma_{nd}$ (where $\sigma_{nd}$ is the inelastic non-diffractive 
cross section) this gives the relation
%
%\begin{equation}
$\sigma_2=\frac{1}{2} \sigma_1^2 / \sigma_{nd}$.
%\label{eq:sig2}
%\end{equation}
%
The experimental groups have used the notation
\begin{equation}
\sigma_2=\frac{1}{2} \frac{\sigma_1^2}{\sigma_{eff}},
\label{eq:sigma2}
\end{equation}
which means that $\sigma_{eff}=\sigma_{nd}$ corresponds to uncorrelated
subcollisions. The experimental results on four-jet events referred to above
find, however, that $\sigma_{eff}$ is much smaller than $\sigma_{nd}$.
Thus at ISR one finds (for jets with $p_\perp > 4$ GeV) $\sigma_{eff} \sim 5$ mb compared to
$\sigma_{nd} \sim 30$ mb, CDF finds for four-jet events 
($p_\perp > 25$ GeV) and 3 jets+$\gamma$ the results $\sigma_{eff} \sim 12$ mb and 
$\sim 14$ mb respectively, to be compared with $\sigma_{nd} \sim 50$ mb.
This means that if there is one subcollision there is an enhanced probability
to have also another one. A possible interpretation is that in central collisions
there are many hard subcollisions, while there are fewer subcollisions in a
peripheral collison.

%\subsection{Pedestal effect}

Another sign of correlations is the observation that in events with a high $p_\perp$ 
jet the underlying event is enhanced, the so called \emph{pedestal effect}. The UA1
collaboration 
at the Sp\={p}S collider studied the $E_\perp$-distribution in $\eta$ around
a jet \cite{Albajar:1988tt}. To avoid the recoiling jet they looked in
$180^\circ$ in azimuth 
on the same side as the jet. The result is that for jets with $E_\perp>5$ GeV
the background level away from the jet is roughly a factor two above the level 
in minimum bias events. Similar results have been observed in resolved
photoproduction by the H1 collaboration \cite{Aid:1995ma}.

\subsection{CDF analysis and the \textsc{Pythia} model}

Rick Field has made very extensive studies of the underlying event at the Tevatron
(see e.g. ref.~\cite{Field:2006iy, *Field:2006ek}).
He has here tuned the \textsc{Pythia} MC to fit CDF data, and found tunes
(e.g. tune A and tune DW) which give very good fits to essentially all data. In
particular he has looked at the $E_\perp$-flow, the charged particle density,
and $p_\perp$-spectra in angular regions perpendicular to a high-$p_\perp$ jet. 
One noticeable result is that the charged multiplicity in this ``transverse'' region grows
rapidly with the $p_\perp$ of the trigger jet up to $p_\perp\!(\mathrm{charged\,\, jet})
\approx 6 \,\mathrm{GeV}$, and then levels off for higher jet energies at twice the
density in minimum
bias events. Also the charged particle spectrum has a much higher tail out to
large $p_\perp$ in events with a high $p_\perp$ jet, compared to the
distribution in minimum bias
events. The multiple collisions have a very important effect in the MC
simulations, and the data cannot be reproduced if
they are not included.

The version of the \textsc{Pythia} MC used by Field is an implementation of an early model
by Sj\"ostrand and van Zijl \cite{Sjostrand:1987su}. In this model it is
assumed that high energy collisions are dominated by hard parton-parton
subcollisions, and also minimum bias events are assumed to have at least one
such subcollision. To be able to reproduce the observed pedestal effect,
the parton distribution is assumed to have a more dense central region, and
is described by a sum of two (three-dimensional) Gaussians. For fixed impact parameter,
$b$, the number of subcollisions is assumed to be given by a Poisson distribution,
with an average proportional to the overlap between the parton distributions in the two
colliding protons. Integrated over the impact parameter this gives a distribution
which actually can be well approximated by a geometric distribution, that is a
distribution with much larger fluctuations than a Poisson. 

The \textsc{Pythia} model does not include diffraction, and describes only
non-diffractive inelastic collisions. Diffraction is related to the
fluctuations via the AGK cutting rules \cite{Abramovsky:1973fm}. In QCD a single pomeron 
exchange can be represented by a gluon ladder. The diagram for double 
pomeron exchange can be cut through zero, one and two of the exchanged pomerons,
with relative weights 1, $-4$, and 2. If we add the contributions to $k$ cut
pomerons from diagrams with an arbitrary number of exchanged pomerons, then we get 
for $k>1$ with the weights in ref \cite{Abramovsky:1973fm} a Poisson distribution.
For fixed impact parameter the assumptions in the \textsc{Pythia} model are thus in agreement
with the AGK rules.

\subsection{Relation $E_\perp - n_{ch}$}

Although Field's tunes of the \textsc{Pythia} model give good fits to data, there
are still problems. The relation between transverse energy and hadron multiplicity
is not what has been expected. In the AGK paper
a cut pomeron was expected to give a chain of hadrons between the remnats of the 
two colliding hadrons, and two cut pomerons should give two such chains and 
therefore doubled particle density. This is in contrast to the CDF data,
where $E_\perp$ grows more than the
multiplicity in multiple collision events. 

The original AGK paper was published
before QCD, and based on a multiperipheral model. However, also in QCD the
DGLAP or BFKL dynamics gives color-connected chains of gluons. In the hadronization
process the gluon exchange ought to give two triplet strings (or cluster chains)
stretching between the projectile remnants, and in the spirit of AGK two cut pomerons
should give four such triplet strings. Field's tunes seem instead to indicate some 
kind of 
color recombination which reduces the effective string length. (Similar recombinations
have been studied by Ingelman and coworkors \cite{Enberg:2001vq}.)

In the \textsc{Pythia} model used by Field different possibilities for the color connection
between the partons involved are studied. The most common parton subcollisions
are $gg \rightarrow gg$, and as mentioned above this is expected to give two strings
between the projectile remnants. Initial state radiation gives extra gluons,
for which the color ordering agrees with the ordering in rapidity. Therefore these
emissions do not increase the total string length very much, and as a
consequence they increase $E_\perp$ more than they increase the hadron multiplicity. 

From the experimental data it was noted
already in ref. \cite{Sjostrand:1987su} that two subcollisions could not
give doubled multiplicity, as expected from four strings as discussed above. 
It was therefore assumed that the second subcollision could give a just single double
string connecting the two outgoing gluons. Another option was replacing the
gluons by a $q\bar{q}$ pair, connected by a single triplet string. This
reduces the multiplicity even further. A third possibility
was to assume that color rearrangment caused the scattered gluons to fit in the
color chains of the first collision, in such a way that the total string length
was increased as little as possile. This gives a minimal additional
multiplicity, and in this case the multiple collisions have an
effect on the total $E_\perp$ and multiplicity similar to the bremsstrahlung gluons
(but the jets are balanced pairwise in transverse momentum).
The default assumption in ref. 
\cite{Sjostrand:1987su} was to give each of these possibilities the same probability, 1/3. In Field's successful tunes these ratios are changed, such that
the last option with color reconnection is chosen in 90\% 
of the cases.

In a more recent study by Sj\"ostrand and Skands \cite{Sjostrand:2004pf}
a number of improvents have been added to the old PYTHIA model. The hope was that 
with these modifications it would be possible to describe data without the
extreme color reconnections which have no real theoretical motivation in QCD. Their result
is, however, discouraging, as they were not able to tune the new model to give
the relation beween $p_\perp$ and multiplicity observed in the data.

\section{Theoretical ideas}

\subsection{Pomeron interactions}

\begin{figure}
%\SetScale{0.5}
\begin{center}
\scalebox{0.31}{\mbox{
 \begin{picture}(420,420)(0,0)
\Line(40,40)(200,40)
\Line(40,400)(200,400)
\Photon(120,40)(120,80){8}{4}
\PhotonArc(143,120)(46,120,240){8}{8}
\PhotonArc(97,120)(46,300,60){8}{8}
\PhotonArc(143,280)(46,120,240){8}{8}
\PhotonArc(97,280)(46,300,60){8}{8}
\PhotonArc(140,235)(57,150,240){7}{6}
\Photon(120,160)(120,240){8}{7}
\Photon(120,320)(120,400){8}{7}
\DashLine(120,160)(120,400){10}
\DashLine(120,40)(120,80){10}
\DashCArc(143,120)(46,120,240){10}
\DashCArc(97,120)(46,300,60){10}
\LongArrow(320,20)(320,425)
\LongArrow(320,20)(380,20)
\Line(360,180)(360,220)
\Line(360,340)(360,380)
\Line(400,100)(400,140)
\CArc(340,20)(20,90,180)
\CArc(380,60)(20,90,180)
\CArc(340,300)(20,90,180)
\CArc(340,60)(20,270,360)
\CArc(340,340)(20,270,360)
\CArc(380,100)(20,270,360)
\CArc(380,140)(20,0,90)
\CArc(340,220)(20,0,90)
\CArc(340,380)(20,0,90)
\CArc(380,180)(20,180,270)
\CArc(340,260)(20,180,270)
\CArc(340,420)(20,180,270)
\Text(305,410)[]{{\Huge $y$}}
\Text(400,20)[l]{{\Huge $\frac{d n}{d y}$}}
\Text(250,4)[]{{\Huge (a)}}
\end{picture}
}}
%\end{center}
%\vspace{-5mm}
\scalebox{1.1}{\mbox{
 \begin{picture}(170,110)(-50,-5)
\Line(5,10)(95,10)
\Line(5,100)(95,100)
\Photon(20,10)(20,100){1}{10}
\Photon(40,10)(40,100){1}{10}
\Photon(60,10)(60,100){1}{10}
\Photon(80,10)(80,100){1}{10}

\Photon(20,20)(40,20){1}{3}
\Photon(20,30)(40,30){1}{3}
\Photon(20,80)(40,80){1}{3}
\Photon(20,90)(40,90){1}{3}
\Photon(60,20)(80,20){1}{3}
\Photon(60,30)(80,30){1}{3}
\Photon(60,80)(80,80){1}{3}
\Photon(60,90)(80,90){1}{3}

\Photon(40,37)(80,37){1}{6}
\Photon(40,51)(80,51){1}{6}
\Photon(40,65)(80,65){1}{6}
\Photon(20,44)(60,44){1}{6}
\Photon(20,58)(60,58){1}{6}
\Photon(20,72)(60,72){1}{6}
\DashLine(50,5)(50,105){4}
\Text(20,105)[]{{\footnotesize 1}} \Text(40,105)[]{{\footnotesize 2}}
\Text(60,105)[]{{\footnotesize 3}} \Text(80,105)[]{{\footnotesize 4}}
\Text(50,-3)[]{{\footnotesize (b)}}
\end{picture}
}}
\end{center}
\caption{\label{fig:pomerons} \textbf{(a)} A pomeron loop can be cut through 0, 1, or 2 of
  its two branches. This can give gaps and bumps in the particle disribution.
  \textbf{(b)} Two pomerons can be represented by four gluons in two color singlet
  pairs. Gluon exchange can switch the pairs (1,2) (3,4) into the singlet
  pairs (1,3) (2,4). A cut as indicated in the figure gives a localized bump
  in the rapidity distribution.}
\end{figure}
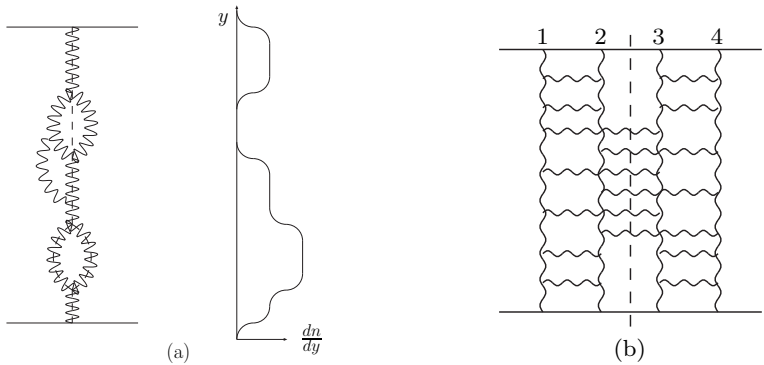

We have to conclude that something important is missing in our understanding
of high energy collisions. 
Although, in the AGK paper, pomeron interactions are assumed to give small 
contributions, pomeron vertices (see e.g. \cite{Bartels:1993ih, *Bartels:1994jj}) 
and pomeron loops may be very important.
As indicated in fig. \ref{fig:pomerons}a, a pomeron loop can give a bump in the 
particle density if both branches of the 
loop are cut, and a gap if the cut passes between the two branches. It is also
conceivable that such gaps and bumps have to be included in a "renormalized"
pomeron \cite{Ostapchenko:2006vr}.

In QCD a pomeron is formed by two 
gluons in a color singlet. Two pomeron exchange thus corresponds to four gluons
in two singlet pairs. If the pairs (1,2) and (3,4) form singlets, then gluon 
exchange can change the
system so that instead the pairs (1,3) and (2,4) form color singlets. This corresponds to
an effective $2 I\!\!P \rightarrow 2 I\!\!P$ coupling
(cf. ref.~\cite{Bartels:2005wa}). A cut with gluons 1 and 2 on one
side and 3 and 4 on the other can then give an isolated bump in the particle
density, as illustrated in fig.~\ref{fig:pomerons}b.
This type of pomeron interactions can also give a bound state 
\cite{Bartels:1993ke, *Bartels:1993it}, which gives a pole
in the angular momentum plane and an essential correction to the normal
cut from the exchange of two uncorrelated pomerons. 

We conclude that there are still many open questions. More experimental information is needed, and
to gain insight into the dynamics it is important to go beyond inclusive observables,
and study observables related to correlations and fluctuations.

\subsection{Dipole cascade models, saturation and pomeron loops}

\begin{figure}
\scalebox{0.55}{\mbox{
\begin{picture}(250,80)(0,-40)
\Vertex(10,80){2}
\Vertex(10,0){2}
\Text(5,80)[]{$\mathrm{x}$}
\Text(5,0)[]{$\mathrm{y}$}
\Line(10,80)(10,0)
\LongArrow(30,40)(60,40)
\Vertex(100,80){2}
\Vertex(100,0){2}
\Vertex(120,50){2}
\Text(95,80)[]{$\mathrm{x}$}
\Text(95,0)[]{$\mathrm{y}$}
\Text(128,50)[]{$\mathrm{z}$}
\Line(100,80)(120,50)
\Line(120,50)(100,0)
\LongArrow(140,40)(170,40)
\Vertex(205,80){2}
\Vertex(225,50){2}
\Vertex(233,30){2}
\Vertex(205,0){2}
\Text(200,80)[]{$\mathrm{x}$}
\Text(200,0)[]{$\mathrm{y}$}
\Text(233,50)[]{$\mathrm{z}$}
\Text(241,30)[]{$\mathrm{w}$}
\Line(205,80)(225,50)
\Line(225,50)(233,30)
\Line(233,30)(205,0)
\Text(125,-37)[]{{\large (a)}}
\end{picture}
}}
%% \begin{center}
\scalebox{0.55}{\mbox{
\begin{picture}(400,175)(0,5)
\Photon(35,130)(60,130){2}{3}
\Photon(365,140)(340,140){2}{3}
\DashLine(60,130)(90,145){2}
\DashLine(60,130)(90,115){2}
\DashLine(340,140)(310,160){2}
\DashLine(340,140)(310,120){2}
\DashLine(90,145)(90,115){2}
\DashLine(310,120)(310,160){2}
\DashLine(90,115)(120,140){2}
\DashLine(120,140)(145,118){2}
\DashLine(160,124)(120,140){2}
\DashLine(310,120)(290,156){2}
\DashLine(310,120)(260,163){2}
\DashLine(260,163)(250,124){2}
\DashLine(250,124)(235,160){2}
\DashLine(235,160)(225,120){2}
\ArrowLine(90,145)(120,140)
\ArrowLine(120,140)(170,165)
\ArrowLine(145,118)(90,115)
\ArrowLine(170,165)(160,124)
\ArrowLine(160,124)(145,118)
\ArrowLine(120,140)(170,165)
\ArrowLine(290,156)(310,160)
\ArrowLine(260,163)(290,156)
\ArrowLine(310,120)(250,124)
\ArrowLine(225,120)(220,161)
\ArrowLine(250,124)(225,120)
\ArrowLine(220,161)(235,160)
\ArrowLine(235,160)(260,163)
\Vertex(310,160){2}
\Vertex(310,120){2}
\Vertex(90,145){2}
\Vertex(90,115){2}
\Vertex(120,140){2}
\Vertex(170,165){2}
\Vertex(145,118){2}
\Vertex(160,124){2}
\Vertex(290,156){2}
\Vertex(260,163){2}
\Vertex(310,120){2}
\Vertex(250,124){2}
\Vertex(235,160){2}
\Vertex(260,163){2}
\Vertex(220,161){2}
\Vertex(225,120){2}
%\Text(25,130)[]{$\gamma^*$}
%\Text(375,140)[]{$\gamma^*$}
\Text(90,155)[]{$q$}
\Text(90,105)[]{$\bar{q}$}
\Text(310,170)[]{$\bar{q}'$}
\Text(310,110)[]{$q'$}
\Text(210,161)[]{$r_3$}
\Text(215,120)[]{$r_4$}
\Text(180,165)[]{$r_1$}
\Text(170,124)[]{$r_2$}
\LongArrow(190,105)(190,75)
\DashLine(170,65)(160,24){2}
\ArrowLine(160,24)(145,18)
\ArrowLine(120,40)(170,65)
\DashLine(225,20)(220,61){2}
\ArrowLine(250,24)(225,20)
\ArrowLine(220,61)(235,60)
\Vertex(220,61){2}
\Vertex(225,20){2}
\Vertex(170,65){2}
\Vertex(160,24){2}
\ArrowLine(170,65)(220,61)
\ArrowLine(225,20)(160,24)
\Text(170,75)[]{$r_1$}
\Text(160,14)[]{$r_2$}
\Text(220,71)[]{$r_3$}
\Text(225,10)[]{$r_4$}
\LongArrow(300,50)(300,75)
\LongArrow(300,50)(325,50)
\Text(300,85)[]{$\mathbf{r}$}
\Text(343,43)[]{$y=$rapidity}
\Text(200,5)[]{{\large (b)}}
\end{picture}
}}
%% \end{center}
\caption{\label{fig:dipev} \textbf{(a)} The evolution of the dipole cascade. 
    At each step, a dipole can split into two new dipoles. 
    \textbf{(b)} A symbolic picture of a $\gamma^* \gamma^*$
    collision in $y-\mathbf{r}_\perp$-space. When two colliding dipoles interact via
    gluon exchange the color connection between the gluons is modified. The
    result is dipole chains stretched between the remnants of the colliding
    systems.}
\end{figure}
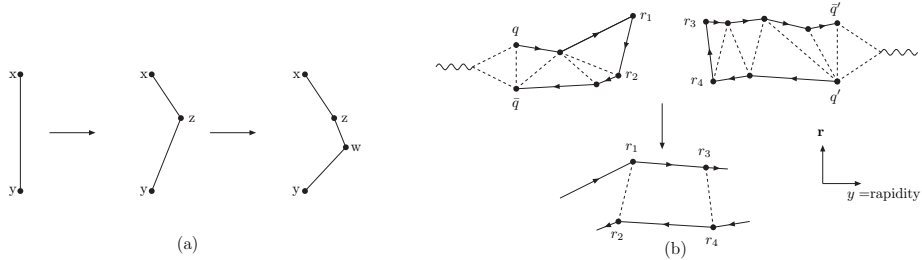
Multiple scattering and rescattering is more easily treated in transverse 
coordinate space. In Mueller's dipole cascade model \cite{Mueller:1993rr,
*Mueller:1994jq, *Mueller:1994gb}
a color dipole formed by
a $q\bar{q}$ pair in a color singlet is split into two dipoles by gluon emission.
Each of these dipoles can split repeatedly into a cascade, see 
fig.~\ref{fig:dipev}a.
The probability per unit rapidity for a split is proportional to $\bar{\alpha}
= N_c \alpha_s/\pi$. When two dipole chains collide, gluon exchange between two 
dipoles implies exchange of color and a recoupling of the chains, as shown in 
fig.~\ref{fig:dipev}b. The probability
for an interaction between two dipoles $i$ and $j$, $f_{ij}$, is proportional to 
$\alpha_s^2 = \pi^2 \bar{\alpha}^2 /N_c^2$, and is thus formally color
suppressed compared to the dipole splitting process.

\begin{figure}
\begin{center}
\epsfig{figure=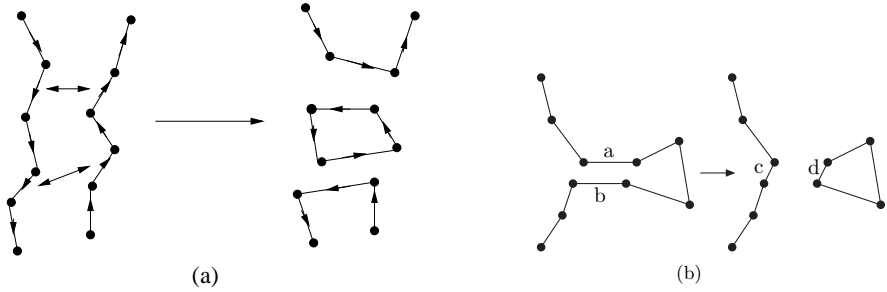,width=5.5cm}
\hspace{10mm}
\scalebox{0.8}{\mbox{
\begin{picture}(180,100)(0,-10)
\Line(10,10)(20,25)
\Line(25,40)(20,25)
\Line(25,40)(50,40)
\Line(50,40)(80,30)
\Line(75,60)(80,30)
\Line(75,60)(55,50)
\Line(30,50)(55,50)
\Line(30,50)(15,70)
\Line(15,70)(10,90)
\LongArrow(85,45)(100,45)
\Line(100,10)(110,25)
\Line(115,40)(110,25)
\Line(115,40)(120,50)
\Line(140,40)(170,30)
\Line(165,60)(170,30)
\Line(165,60)(145,50)
\Line(145,50)(140,40)
\Line(120,50)(105,70)
\Line(105,70)(100,90)
\Vertex(10,10){2}
\Vertex(20,25){2}
\Vertex(25,40){2} 
\Vertex(50,40){2} 
\Vertex(80,30){2} 
\Vertex(75,60){2}
\Vertex(55,50){2} \Vertex(30,50){2} \Vertex(15,70){2} \Vertex(10,90){2}
\Vertex(100,10){2} \Vertex(110,25){2} \Vertex(115,40){2} \Vertex(120,50){2}
\Vertex(140,40){2} \Vertex(170,30){2} \Vertex(165,60){2} \Vertex(145,50){2}
\Vertex(105,70){2} \Vertex(100,90){2}
\Text(38,35)[]{b} \Text(42,55)[]{a} \Text(113,47)[]{c} \Text(139,48)[]{d} 
\Text(80,-3)[]{{\small (b)}}
\end{picture}
}}
\end{center}
\caption{\label{fig:multichain}  \textbf{(a)} If more than one pair of dipoles 
  interact it can result in dipole loops, which correspond to pomeron loops.
  \textbf{(b)} Schematic picture of a dipole swing. 
  If the two dipoles $a$ and $b$ have the same color, they can be replaced
  by the dipoles $c$ and $d$. The result is a closed loop formed within an 
  individual dipole cascade.}  
\end{figure}

In the eikonal approximation the total scattering probability is determined by
the expression $1 - \prod_{ij}(1-f_{ij})$, which is always smaller than 1 and thus
satisfies the constraints from unitarity. As seen in fig.~\ref{fig:multichain}a,
multiple dipole-dipole interactions can imply that the color dipoles form
closed loops, which correspond to the pomeron loops in fig.~\ref{fig:pomerons}a.

Mueller's model includes those pomeron loops, which correspond to cuts in the particular 
Lorentz frame used for the calculation, but not loops which are fully inside
one of the colliding cascades. This implies that the formalism is not Lorentz
frame independent, and different ways have been suggested to achieve
a frame independent formulation (see e.g. refs.~\cite{Iancu:2005dx, Kozlov:2007xc}). 
However, so far no explicitely frame independent
formalism has been presented.  

In one approach the evolution is expressed in terms of \emph{interacting} dipoles.
This implies that the number of dipoles can be reduced, and the evolution of the
projectile cascade depends on the target. Besides the $1\rightarrow 2$
dipole vertex there should here also be a $2\rightarrow 1$ vertex.
In another approach the evolution of the projectile is independent of the target,
and the non-interacting dipoles are eliminated afterwards. In this approach there 
is no need to reduce the number of dipoles in the evolution.

\emph{Dipole swing}

A model based on the latter approach is presented in ref. \cite{Avsar:2006jy}.
In this model pomeron loops can be formed with the help of a 
recoupling of the dipole chains, a "dipole swing". Just as the dipole-dipole
scattering, the pomeron loops in the cascades should be color suppressed.
With a finite number of colors we can have not only dipoles but also higher
color multipoles. Two charges and two anticharges with the same color may 
be better approximated by two dipoles formed by nearby charge-anticharge pairs.
These pairs may be different from the initially generated dipoles, and
the result is a recoupling of the dipole chain, as seen in
fig. \ref{fig:multichain}b. The same effect can 
also be obtained from gluon exchange, which is 
proportional to $\alpha_s$ and thus also color suppressed cf. to the dipole
splitting proportional to $\bar{\alpha}$.

\begin{figure}
  \includegraphics[angle=270,  scale=0.41]{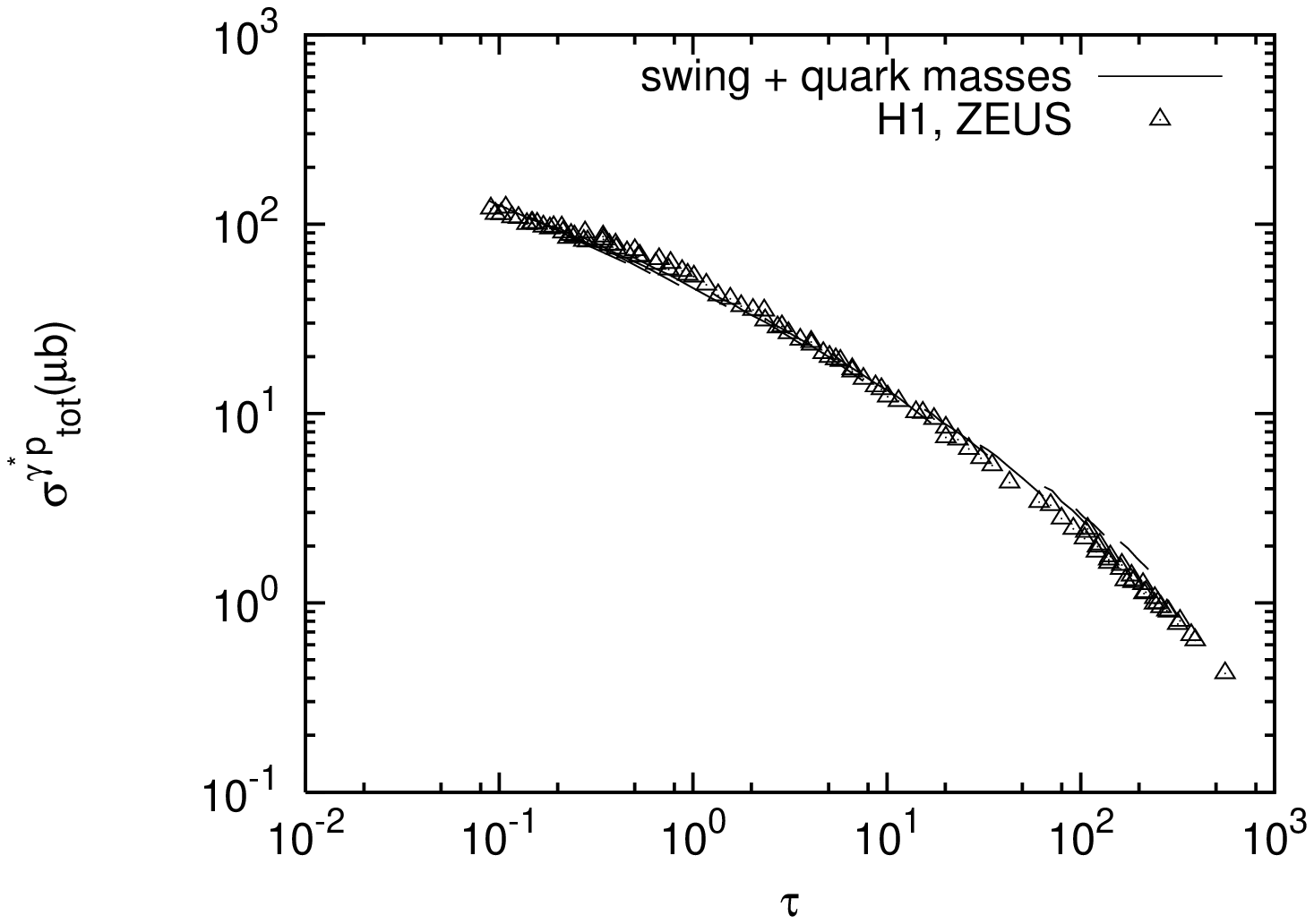}%
  \includegraphics[angle=270, scale=0.49]{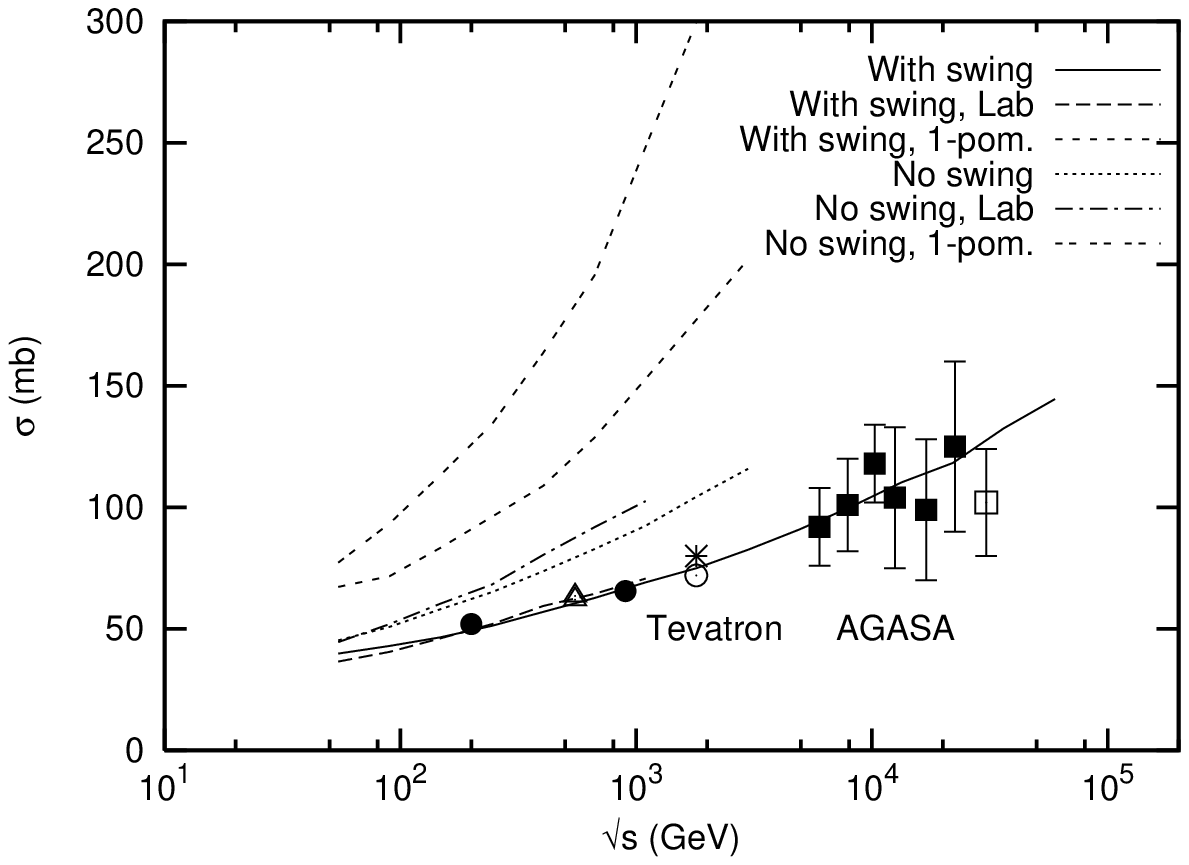}
  \caption{\emph{Left}: The $\gamma p$ total cross section obtained from the model in
 %    ref.~\cite{Avsar:2006jy, Avsar:2007ht} cf. to data from H1 and Zeus.
     The result is plotted as a 
     function of the scaling variable $\tau=Q^2/Q_{s}^2$, where $Q_{s}^2=
     Q_0^2 (x_0/x)^\lambda$ with $Q_0=1\mathrm{GeV}$, $x_0=3\cdot 10^{-4}$, 
     $\lambda=0.29$. 
   \emph{Right}: The total $pp$ scattering cross section. Results are presented
    for evolution with and without the dipole swing mechanism. The
    one pomeron result and the
    result obtained in a frame where one of the protons is almost at rest
    are also shown.}
  %  Data points are taken from \cite{Adloff:2000qk} and \cite{Breitweg:2000yn}.
    \label{fig:results} 
\end{figure}

The swing does not result in a reduction of the number of dipoles, but
the saturation effect is obtained as the recoupled dipoles are smaller and
therefore have smaller cross sections. Inserted in a MC the result is 
approximately frame independent, and the model describes well both
the $F_2$ structure function in DIS and the $pp$ scattering cross section
\cite{Avsar:2006jy, Avsar:2007ht}, as shown in fig.~\ref{fig:results}.
(For these results also energy conservation and a running $\alpha_s$ are very
important \cite{Avsar:2005iz}.)
We see here that the $\gamma^* p$ cross section satisfies geometric
scaling. The $pp$ cross section is reduced by about a factor 4 cf. to
the one pomeron exchange at the Tevatron, and we also see that the result of the model is
the same when calculated in the cms as in the rest frame of the target proton,
if pomeron loops are included also in the evolution via the dipole swing.

Besides the total cross sections it is also possible to calculate the probability
to have pomeron loops formed by multiple collisions in a given frame, or
loops formed within the cascades. As examples we find at the Tevatron in the cms
on average 2.2 loops from multiple collisions and 0.65 loops in each of the two
cascades. In an asymmetric frame, where the total rapidity range is devided in
4.5 + 10.5 units, we find instead 2 loops from multiple collisions,
and 0.15 and 1.35 in the two cascades respectively. In both cases this gives in 
total 3.5 loops. At LHC we obtain in the same 
way in total an average of 5 loops.

Using the eikonal approximation it is besides total cross sections also possible to calculate elastic scattering and diffractive excitation \cite{Avsar:2007b}, but so far it has 
not been possible to calculate exclusive final states. The aim for the future is to 
bridge the gap between dipole cascades, AGK, and traditional MC generators,
and construct an event generator fully compatible with unitarity and the AGK
cutting rules.

\section{Conclusions}
\label{conclusions}

\vspace{2mm}
\begin{itemize}
\item{Multiple collisions are present in data.}
\item{Hard subcollisions are correlated. The underlying event is different from
a minimum bias event.}
\item{Rick Field's tunes of the PYTHIA MC fit Tevatron data well, but the relation between transverse energy and multiplicity is not understood.
This may indicate some kind of color rearrangement, or a "renormalized pomeron".}
\item{Multiple collisions and unitarity constraints are easier treated in 
transverse coordinate space. The dipole formalism can describe $I\!\!P$ loops
and diffraction.
The application of AGK cutting rules then imply the presence of rapidity gaps.}
\item{For the future we hope to be able to combine the dipole formalism and traditional
MC generators to obtain event generators which include diffraction and are
compatible with unitarity and AGK.}
\end{itemize}

%------------------------------------------------------------------------------
%       Bibliography
%------------------------------------------------------------------------------
\begin{footnotesize}
\bibliographystyle{blois07} 
{\raggedright
\bibliography{blois07g}

\providecommand{\etal}{et al.\xspace}
\providecommand{\href}[2]{#2}
\providecommand{\coll}{Coll.}
\catcode`\@=11
\def\@bibitem#1{%
\ifmc@bstsupport
  \mc@iftail{#1}%
    {;\newline\ignorespaces}%
    {\ifmc@first\else.\fi\orig@bibitem{#1}}
  \mc@firstfalse
\else
  \mc@iftail{#1}%
    {\ignorespaces}%
    {\orig@bibitem{#1}}%
\fi}%
\catcode`\@=12
\begin{mcbibliography}{10}

\bibitem{Sjostrand:2004pf}
T.~Sjöstrand and P.~Z. Skands,
\newblock JHEP{} {\bf 03},~053~(2004).
\newblock \href{http://www.arXiv.org/abs/hep-ph/0402078}{{\tt
  hep-ph/0402078}}\relax
\relax
\bibitem{Sjostrand:1987su}
T.~Sjöstrand and M.~van Zijl,
\newblock Phys. Rev.{} {\bf D36},~2019~(1987)\relax
\relax
\bibitem{Gustafson:1999kh}
G.~Gustafson and G.~Miu,
\newblock Phys. Rev.{} {\bf D63},~034004~(2001).
\newblock \href{http://www.arXiv.org/abs/hep-ph/0002278}{{\tt
  hep-ph/0002278}}\relax
\relax
\bibitem{Gustafson:2002kz}
G.~Gustafson, L.~Lönnblad, and G.~Miu,
\newblock Phys. Rev.{} {\bf D67},~034020~(2003).
\newblock \href{http://www.arXiv.org/abs/hep-ph/0209186}{{\tt
  hep-ph/0209186}}\relax
\relax
\bibitem{Cline:1973kv}
D.~Cline, F.~Halzen, and J.~Luthe,
\newblock Phys. Rev. Lett.{} {\bf 31},~491~(1973)\relax
\relax
\bibitem{Ellis:1973nb}
S.~D. Ellis and M.~B. Kislinger,
\newblock Phys. Rev.{} {\bf D9},~2027~(1974)\relax
\relax
\bibitem{Durand:1987yv}
L.~Durand and P.~Hong,
\newblock Phys. Rev. Lett.{} {\bf 58},~303~(1987)\relax
\relax
\bibitem{Akesson:1986iv}
{ Axial Field Spectrometer} Collaboration, T.~Akesson {\em et al.},
\newblock Z. Phys.{} {\bf C34},~163~(1987)\relax
\relax
\bibitem{Abe:1993rv}
{ CDF} Collaboration, F.~Abe {\em et al.},
\newblock Phys. Rev.{} {\bf D47},~4857~(1993)\relax
\relax
\bibitem{Abazov:2002mr}
{ D0} Collaboration, V.~M. Abazov {\em et al.},
\newblock Phys. Rev.{} {\bf D67},~052001~(2003).
\newblock \href{http://www.arXiv.org/abs/hep-ex/0207046}{{\tt
  hep-ex/0207046}}\relax
\relax
\bibitem{Abe:1997xk}
{ CDF} Collaboration, F.~Abe {\em et al.},
\newblock Phys. Rev.{} {\bf D56},~3811~(1997)\relax
\relax
\bibitem{Gwenlan:2002st}
{ ZEUS} Collaboration, C.~Gwenlan,
\newblock Acta Phys. Polon.{} {\bf B33},~3123~(2002)\relax
\relax
\bibitem{Albajar:1988tt}
{ UA1} Collaboration, C.~Albajar {\em et al.},
\newblock Nucl. Phys.{} {\bf B309},~405~(1988)\relax
\relax
\bibitem{Aid:1995ma}
{ H1} Collaboration, S.~Aid {\em et al.},
\newblock Z. Phys.{} {\bf C70},~17~(1996).
\newblock \href{http://www.arXiv.org/abs/hep-ex/9511012}{{\tt
  hep-ex/9511012}}\relax
\relax
\bibitem{Field:2006iy}
{ CDF} Collaboration, R.~D. Field.
\newblock Presented at 33rd International Conference on High Energy Physics
  (ICHEP 06), Moscow, Russia, 26 Jul - 2 Aug 2006\relax
\relax
\bibitem{Field:2006ek}
{ CDF} Collaboration, R.~Field.
\newblock Presented at 34th International Meeting on Fundamental Physics: From
  HERA and the Tevatron to the LHC, El Escorial, Madrid, Spain, 2-7 Apr
  2006\relax
\relax
\bibitem{Abramovsky:1973fm}
V.~A. Abramovsky, V.~N. Gribov, and O.~V. Kancheli,
\newblock Yad. Fiz.{} {\bf 18},~595~(1973)\relax
\relax
\bibitem{Enberg:2001vq}
R.~Enberg, G.~Ingelman, and N.~Timneanu,
\newblock Phys. Rev.{} {\bf D64},~114015~(2001).
\newblock \href{http://www.arXiv.org/abs/hep-ph/0106246}{{\tt
  hep-ph/0106246}}\relax
\relax
\bibitem{Bartels:1993ih}
J.~Bartels,
\newblock Z. Phys.{} {\bf C60},~471~(1993)\relax
\relax
\bibitem{Bartels:1994jj}
J.~Bartels and M.~Wusthoff,
\newblock Z. Phys.{} {\bf C66},~157~(1995)\relax
\relax
\bibitem{Ostapchenko:2006vr}
S.~Ostapchenko,
\newblock Phys. Lett.{} {\bf B636},~40~(2006).
\newblock \href{http://www.arXiv.org/abs/hep-ph/0602139}{{\tt
  hep-ph/0602139}}\relax
\relax
\bibitem{Bartels:2005wa}
J.~Bartels, M.~Salvadore, and G.~P. Vacca,
\newblock Eur. Phys. J.{} {\bf C42},~53~(2005).
\newblock \href{http://www.arXiv.org/abs/hep-ph/0503049}{{\tt
  hep-ph/0503049}}\relax
\relax
\bibitem{Bartels:1993ke}
J.~Bartels and M.~G. Ryskin,
\newblock Z. Phys.{} {\bf C60},~751~(1993)\relax
\relax
\bibitem{Bartels:1993it}
J.~Bartels and M.~G. Ryskin,
\newblock Z. Phys.{} {\bf C62},~425~(1994)\relax
\relax
\bibitem{Mueller:1993rr}
A.~H. Mueller,
\newblock Nucl. Phys.{} {\bf B415},~373~(1994)\relax
\relax
\bibitem{Mueller:1994jq}
A.~H. Mueller and B.~Patel,
\newblock Nucl. Phys.{} {\bf B425},~471~(1994).
\newblock \href{http://www.arXiv.org/abs/hep-ph/9403256}{{\tt
  hep-ph/9403256}}\relax
\relax
\bibitem{Mueller:1994gb}
A.~H. Mueller,
\newblock Nucl. Phys.{} {\bf B437},~107~(1995).
\newblock \href{http://www.arXiv.org/abs/hep-ph/9408245}{{\tt
  hep-ph/9408245}}\relax
\relax
\bibitem{Iancu:2005dx}
E.~Iancu, G.~Soyez, and D.~N. Triantafyllopoulos,
\newblock Nucl. Phys.{} {\bf A768},~194~(2006).
\newblock \href{http://www.arXiv.org/abs/hep-ph/0510094}{{\tt
  hep-ph/0510094}}\relax
\relax
\bibitem{Kozlov:2007xc}
M.~Kozlov, E.~Levin, and A.~Prygarin,
\newblock Nucl. Phys.{} {\bf A792},~122~(2007).
\newblock \href{http://www.arXiv.org/abs/arXiv:0704.2124 [hep-ph]}{{\tt
  arXiv:0704.2124 [hep-ph]}}\relax
\relax
\bibitem{Avsar:2006jy}
E.~Avsar, G.~Gustafson, and L.~Lönnblad,
\newblock JHEP{} {\bf 01},~012~(2007).
\newblock \href{http://www.arXiv.org/abs/hep-ph/0610157}{{\tt
  hep-ph/0610157}}\relax
\relax
\bibitem{Avsar:2007ht}
E.~Avsar and G.~Gustafson,
\newblock JHEP{} {\bf 04},~067~(2007).
\newblock \href{http://www.arXiv.org/abs/hep-ph/0702087}{{\tt
  hep-ph/0702087}}\relax
\relax
\bibitem{Avsar:2005iz}
E.~Avsar, G.~Gustafson, and L.~Lönnblad,
\newblock JHEP{} {\bf 07},~062~(2005).
\newblock \href{http://www.arXiv.org/abs/hep-ph/0503181}{{\tt
  hep-ph/0503181}}\relax
\relax
\bibitem{Avsar:2007b}
E.~Avsar, G.~Gustafson, and L.~Lönnblad.
\newblock Preprint in preparation\relax
\relax
\end{mcbibliography}
}
\end{footnotesize}
\end{document}